\documentclass[3p,authoryear,preprint,12pt]{elsarticle}

\usepackage{graphicx}
\usepackage{tabulary}
\usepackage{epsfig}
\usepackage{amssymb}
\usepackage{multirow}
\usepackage{appendix}
\usepackage{natbib}
\usepackage{lineno}
\newcommand{\ch}{Chelyabinsk}

\newcommand{\beq}[1]{\begin{equation}\label{#1}}
\newcommand{\eeq}{\end{equation}}

\journal{\tt arXiv}
\begin{document}

%%%%%%%%%%%%%%%%%%%%%%%%%%%%%%%%%%%%%%%%%%%%%%%%%%%%%%%%%%%%%%%%%%%%%%%%%%%%%%%%%
%FRONT MATTER
%%%%%%%%%%%%%%%%%%%%%%%%%%%%%%%%%%%%%%%%%%%%%%%%%%%%%%%%%%%%%%%%%%%%%%%%%%%%%%%%%
\begin{frontmatter}

\title{A preliminary reconstruction of the orbit of the Chelyabinsk
  Meteoroid}

\author{Jorge I. Zuluaga}
\ead{jzuluaga@fisica.udea.edu.co}
\author{Ignacio Ferrin}
\ead{respuestas2013@gmail.com}

\address{Instituto de F\'isica - FCEN, Universidad de
  Antioquia,\\ Calle 67 No. 53-108, Medell\'in, Colombia}

%%%%%%%%%%%%%%%%%%%%%%%%%%%%%%%%%%%%%%%%%%%%%%%%%%%%%%%%%%%%%%%%%%%%%%%%%%%%%%%%%
%ABSTRACT
%%%%%%%%%%%%%%%%%%%%%%%%%%%%%%%%%%%%%%%%%%%%%%%%%%%%%%%%%%%%%%%%%%%%%%%%%%%%%%%%%
\begin{abstract}
In February 15 2013 a medium-sized meteoroid impacted the atmosphere
in the region of Chelyabinsk, Russia.  After its entrance to the
atmosphere and after travel by several hundred of kilometers the body
exploded in a powerful event responsible for physical damages and
injured people spread over a region enclosing several large cities.
We present in this letter the results of a preliminary reconstruction
of the orbit of the {\it Chelyabinsk meteoroid}.  Using evidence
gathered by one camera at the Revolution Square in the city of
Chelyabinsk and other videos recorded by witnesses in the close city
of Korkino, we calculate the trajectory of the body in the atmosphere
and use it to reconstruct the orbit in space of the meteoroid previous
to the violent encounter with our planet.  In order to account for the
uncertainties implicit in the determination of the trajectory of the
body in the atmosphere, we use Monte Carlo methods to calculate the
most probable orbital parameters and their dispersion.  Although the
orbital elements are affected by uncertainties the orbit has been
sucesfully reconstructed.  We use it to classify the meteoroid among
the near Earth asteroid families finding that the parent body belonged
to the Apollo asteroids.
\vspace{0.3cm}
\end{abstract}

%%%%%%%%%%%%%%%%%%%%%%%%%%%%%%%%%%%%%%%%%%%%%%%%%%%%%%%%%%%%%%%%%%%%%%%%%%%%%%%%%
%KEY WORDS
%%%%%%%%%%%%%%%%%%%%%%%%%%%%%%%%%%%%%%%%%%%%%%%%%%%%%%%%%%%%%%%%%%%%%%%%%%%%%%%%%

\begin{keyword}
Meteors\sep Asteroid, dynamics
\end{keyword}

\end{frontmatter}

%%%%%%%%%%%%%%%%%%%%%%%%%%%%%%%%%%%%%%%%%%%%%%%%%%%%%%%%%%%%%%%%%%%%%%%%%%%%%%%%%
%PAPER CONTENT
%%%%%%%%%%%%%%%%%%%%%%%%%%%%%%%%%%%%%%%%%%%%%%%%%%%%%%%%%%%%%%%%%%%%%%%%%%%%%%%%%

%%%%%%%%%%%%%%%%%%%%%%%%%%%%%%%%%%%%%%%%%%%%%%%%%%%%%%%%%%%%%%%%%%%%%%%%%
\section{Introduction}
\label{sec:introduction}
%%%%%%%%%%%%%%%%%%%%%%%%%%%%%%%%%%%%%%%%%%%%%%%%%%%%%%%%%%%%%%%%%%%%%%%%%

In February 15 2013 at about 09:20 local time a large fireball
followed by a huge explosion was seen in the skies of the Chelyabinsk
region in Russia.  Cameras all across the region registered the
historic event creating an ah-hoc network of instruments able to
potentially provide enough visual information for a reconstruction of
the trajectory of the body in the atmosphere and from it the orbit of
the meteoroid.

The ballistic reconstruction of the orbit of bodies associated to
bright fireballs is one of the tools used to study the nature and
origin of these bodies.  There are a relatively small number of
bright-bolides whose orbit has been succesfully reconstructed, even
despite the relatively large number of cases reported every year (see
e.g. \citealt{Trigo2009}).  In some cases the lack of enough number of
observations or a limited quality in the available observations
renders impossible the reconstruction of the orbit.

In the case of the Chelyabinsk meteoroid the number of observations
  and the quality of some of them seem to be enough for a successful
reconstruction of the meteoroid orbit.  More than several tens of
videos, ranging from amateur videos, videos recorded by onboard
vehicle cameras and cameras of the public transit and police network
are readily available in the web (for a large although incomplete
compilation of videos see {\tt http://goo.gl/rbdZm}\footnote{Hereafter
  and to save space we will shorten all urls.})

In this letter we present one of the first rigorous attempts to
reconstruct the orbit of the Chelyabinsk Meteoroid. We use here the
recording of a camera located in the Revolutionary Square in
Chelyabinsk and one video recorded in the close city of Korkino.  Both
observations are used to triangulate the trajectory of the body in the
atmosphere.  The method used is here was first devised by Stefen Geen
and published in one his blog, {\it Ogle Earth} {\tt
  http://goo.gl/vcG3Y} in February 16 2013.  

Further details, updates, videos and additional images and plots that
those released on this letter are available at {\tt
  http://astronomia.udea.edu.co/chelyabinsk-meteoroid}.

Although an analysis of the data taken by scientific instruments in
the affected area, combined with further analyses of the abundant
information gathered by eye witnesses and in situ cameras, would allow
a precise determination of the meteoroid orbit, a first attempt to
reconstruct the orbit of the meteoroid would help us to ellucidate its
nature and origin and would certainly improve the on-going research on
the event.

%%%%%%%%%%%%%%%%%%%%%%%%%%%%%%%%%%%%%%%%%%%%%%%%%%%%%%%%%%%%%%%%%%%%%%%%%
\section{Trajectory in the atmosphere}
\label{sec:trajectory}
%%%%%%%%%%%%%%%%%%%%%%%%%%%%%%%%%%%%%%%%%%%%%%%%%%%%%%%%%%%%%%%%%%%%%%%%%

To reconstruct the trajectory in the atmosphere we use the same method
and images originally used by Stefan Geen and publickly available in
his blog {\it Ogle Earth}.  The original blog entry is available here
{\tt http://goo.gl/vcG3Y}. It is interesting to stress that at using
the methods and results published in the Geen's blog, we are
recognizing the fundamental contribution that enthusiastic people
would have in specific scientific achievements.  Similar cases of
interaction between active enthusiastic contributors (a.k.a. {\it
  citizen astronomers}) and professionals have been recently seen in
other areas in astronomy (see e.g. \citealt{Lintott2008} and
\citealt{Fischer2012})

The method cleverly devised by Geen uses the shadow cast by light
poles at the Revolution Square of Chelyabinsk during the flyby of the
fireball, to estimate the elevation and azimuth of the meteoroid at
different stages if its impact with the atmosphere (the original video
can be found at {\it http://goo.gl/nNcvq}).  To calculate elevations
he height of the light poles is estimated by comparing them with their
separation as given by Google Earth images and tools.  Azimuts are
calculated from the angle subtended by the shadows and the border of
the Prospect Lenina street in front of the square.  The street and the
square are almost perfectly aligned with the East-West and South-North
directions.

For our reconstruction of the trajectory we have selected two
particular times in the images recorded by a public camera located in
front of the square.  The first one correspond to the point when the
fireball reaches a brightness enough to give clear shadows of the
poles.  We call this point the "brightening point" (BP).  A local
elevation $a_{\rm BP}=33^o$ and an azimut $A_{\rm EP}=122^o$ were
found after analysing the corresponding video frame.  The second time
coincides aproximately with the closest point of the trajectory to the
observing point.  It also coincides with the time where the bolid
starts to fragmentate.  We call this second reference point the
``fragmentation point'' (FP).  The FP elevation and azimut are
measured as $h_{\rm FP}=32^o$ and $A_{\rm EP}=222^o$.  The time
required by the meteoroid to travel between those two points is
estimated from the time stamps in the video and it was found equal to
$\Delta t=3.5$ s.  We assume that the local time of the meteoroid
passage by BP is precisely provided by the time stamps in the
video. Hereafter we will assume $t_{\rm BP}=$9 h 20 m 29 s.  This time
could be affected by uncertainties up to few seconds.

The information provided by a single observer is not enough to
determine the meteoroid trajectory in the atmosphere.  In particular
we need an independent observation to estimate for example the
distance and height of BP and FP with respect to the Revolution
Square.  There are several tens of independent and publickly available
videos shoot at very different points in a radius of almost 700 km.
However most of them were recorded by amateur equipment or cameras in
vehicles making complicated to find a suitable parable of the
``virtual observations'' performed at Revolution Square.  At the time
of writing of this letter no video was yet identified with the
required information to triangulate in a precise way the meteoroid
trajectory.  

There is an extraordinary coincidence that help us to constraint the
distance of the meteoroid to the center of Cheliabinsk.  Eye witnesses
and videos show that at Korkin a small city to the south of
Chelyabinsk, the fireball streak across the local zenith.  In the
video publickly available here {\tt http://goo.gl/0HhDR} it can be
seen how the meteoroid was moving close to the zenith when the body
exploded above the place were the video was shot.  The precise
location of the recording camera has not been yet reported by the
author.  Other observers at Korkino registered in video the huge
contrail left by the meteoroid.  Their observations confirm that the
fireball passed almost through the first vertical.  In those cases
again no precise information of the geographic location of the camara
was either provided by the authors.

Finally a third vantage point should be selected to complete the
triangulation of the meteoroid.  As originally suggested by Stefan
Geen the third point was chosen to be the surface of Lake Chebarkul,
one of the many lakes in the Chelyabinsk region and so far the only
place where a large hole in the ice were preliminarly associated with
the impact of a large fragment of the meteoroid.  Assuming that the
fragment traveled in the same direction that the meteoroid in the
atmosphere, Lake Chebarkul provie the intersection of the meteoroid
trajaectory with the Earth surface.

\begin{figure}  
  \centering
   \includegraphics[width=0.9
  \textwidth]{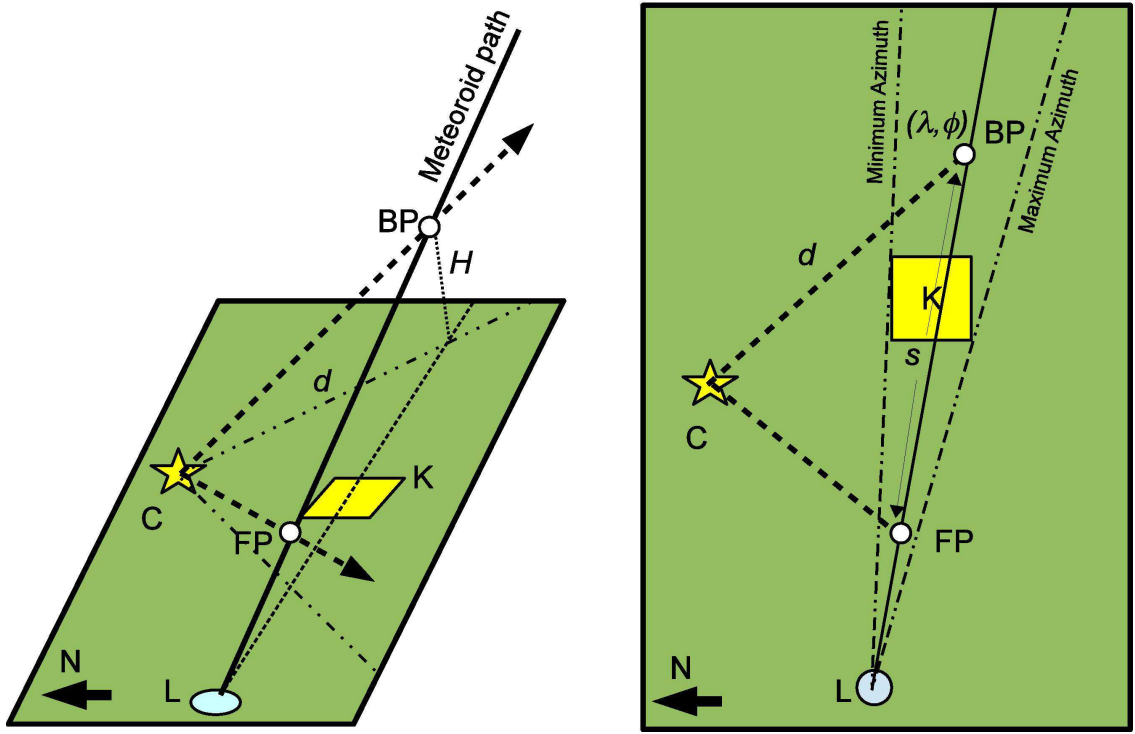}
  \scriptsize
  \caption{ Schematic representation of the trajectory of the
    meteoroid with respect to three vantages points used here to
    triangulate the meteoroid trajectory: the Revolution Square at
    Chelyabinsk (C), the Korkin metropolitan area (K) and Lake
    Chebarkul (L).  The brightening point (BP) and the fragmentation
    point (FP) are the points of the trajectory seen from Chelyabinsk
    and measured using the shadow cast by the poles at the
    Revolutionary Square.  Observations at K constraint the azimuth of
    the trajectory (solid line in the right panel) to be between a
    minimum and a maximum azimuth.  At a given azimuth the distance
    $d$ between C and the surface point below BP, the height $H$ of
    BP, and the distance $s$ traveled by the meteoroid between BP and
    FP, can be calculated.  The latitude ($\phi$) and longitude
    ($\lambda$) of the BP were assumed as the planetocentric spherical
    coordinates of the initial point of the reconstructed orbit of the
    meteoroid.
    \vspace{0.2cm}} 
  \label{fig:trajectory}
\end{figure}

Combining the information provided by the shadows at the center of
Chelyabinsk, the videos and observations of the fireball and its
contrail at Gorkin and the impact at lake Chebarkul, the trajectory of
the meteoroid in the atmosphere can be constrained to the region
depicted in Figure \ref{fig:trajectory}.

There are six critical properties describing the trajectory in the
atmosphere and that we need to estimate in order to proceed at
reconstructing the orbit in space.  The linear height $H$ of the
reference point BP; the elevation $a$ and azimuth $A$ of the meteor
radiant which is the same elevation and aizmut of the trajectory as
seen from the impact point; the latitude $\phi$, and longitude
$\lambda$ of the surface point right below BP, and the velocity $v$ of
the meteoroid at SP that we will assume equal to its orbital velocity.

All these properties are function of a single unknown free parameter:
the distance $d$ between the central square at \ch (C) and the surface
point below BP (see right panel in figure \ref{fig:trajectory}.  Each
value of $d$ correspond to a value of the azimut of the trajectory
compatible with the fact that the fireball streaks the sky close or at
the zenith at some points in Korkino.  The latter condition implies
that $d$ is between 50 and 72 km.

In table 1 we present the properties of the trajectories defined by
the extreme values of the independent parameter $d$.

%TTTTTTTTTTTTTTTTTTTTTTTTTTTTTTTTTTTTTTTTTTTTTTTTTTTTTTTTTTTTTTTTTTTTTTTTTTT
%TABLE 1: TRAJECTORY PROPERTIES
\begin{table}[ht]
  \centering
  \begin{tabular}{llccl}
    \hline\hline
    Property & Symbol & $d$ = 50 km & $d$ = 72 km & Units\\\hline
    Height at BP & $H_{\rm BP}$ & 32.47 & 46.75 & km\\
    Elevation BP & $h$ & 16.32 & 19.73 & degree\\
    \bf Azimut BP & $A$ & \bf 91.60 & \bf 96.48 & degree\\
    Latitude below BP & $\phi$ & 54.92 & 54.81 & degree\\
    Longitude below BP & $\lambda$ & 62.06 & 62.35 & degree\\
    Height at FP & $H_{\rm FP}$ & 20.31 & 25.04 & km\\
    Radiant declination & $\delta$ & 12.38 & 12.39 & degree\\
    Radiant right ascension & RA & 22.44 & 22.07 & hour\\
    Meteoroid velocity & $v$ & 13.43 & 19.65 & km/s\\
    \hline\hline
  \end{tabular}
  \caption{Properties of the trajectory for two extreme values of the
    horizontal distance $d$ between the Revolutionary Square at
    Chelyabinsk and the meteoroid brightening point (BP).  The
    equatorial coordinates of the radiant $\delta$ and RA were
    calculated assuming that the meteoroid was at BP at 9 h 20 m 29 s.
    Heights and velocity are measured with respect to the surface of
    the Earth.
  \label{tab:trajectory}}
\end{table}
%TTTTTTTTTTTTTTTTTTTTTTTTTTTTTTTTTTTTTTTTTTTTTTTTTTTTTTTTTTTTTTTTTTTTTTTTTTT

According to our estimations, the Chelyabinski meteor started to
brighten up when it was between 32 and 47 km up in the atmosphere.
The radiant of the meteoroid was located in the constellation of
Pegasus (northern hemisphere).  At the time of the event the radiant
was close to the East horizon where the sun was starting to rise (this
is confirmed by many videos showing the first appearance of the meteor
during the twilight, see an example at {\tt }).  The velocity of the
body predicted by our analysis was between 13 and 19 km/s (relative to
the Earth) which encloses the preferred figure of 18 km/s assumed by
other researchers. The relatively large range of velocities compatible
with our uncertainties in the direction of the trajectory, represent
the largest source of dispersion in the reconstruction of the orbit.

%%%%%%%%%%%%%%%%%%%%%%%%%%%%%%%%%%%%%%%%%%%%%%%%%%%%%%%%%%%%%%%%%%%%%%%%%
\section{Orbit reconstruction}
\label{sec:orbit}
%%%%%%%%%%%%%%%%%%%%%%%%%%%%%%%%%%%%%%%%%%%%%%%%%%%%%%%%%%%%%%%%%%%%%%%%%

Once the basic properties of the trajectory in the atmosphere have
been estimated we need to select a position of the body in the
trajectory and express it with respect to a local reference frame.
The selected position will be the initial point of our reconstructed
orbit.

The position of the initial point is expressed first in spherical
coordinates (latitude, longitude and height with respect to the
geoid).  We need to transform it first to the International
Terrestrial Reference System (ITRS) expressing it in cartesian
coordinates.  Then the position should be transformed to the
Geocentric Celestial Reference System (GCRS) in order to account for
the rotation of the Earth.  Finally and before we proceed with the
integration, we need to rotate the position to a planetocentric
ecliptic coordinate system.

We perform all these tasks using NOVAS, the Naval Observatory Vector
Astrometry Software developed and distributed by the U.S. Navy
Observatory (USNO) \citep{Bangert2011}\footnote{The package and its
  documentation can be download from {\tt http://goo.gl/58kLt}}.  In
all cases we have checked that the numbers returned in the
transformation procedures are in agreement with what should be
expected.

In order to integrate the orbit we construct a gravitational scenario
including the 8 major solar system planets plus the Earth's Moon.  We
compute the exact position of these bodies at the precise time of the
impact using the JPL DE421 ephemeris kernel and the NAIF/SPICE toolkit
\citep{Acton1996}
\footnote{URL: {\tt http://naif.jpl.nasa.gov/naif/toolkit.html}.}

To integrate the orbit we use Mercury \citep{Chambers2008}.  Mercury
allows us to integrate backwards the trajectory of the meteoroid
taking into account potential perturbing close encounters with the
Earth, Moon or Mars.  In all cases the orbit of the meteoroid was
integrated backwards, 4 years before the impact.

In order to account for the uncertainties in the trajectory we
integrate the orbit of 50 different bodies having initial point BP
with properties compatible with the uncertainties described before.
To perform this integration we generated 50 random values of the
distance $d$ uniformly distributed between 50 and 72 km.  The rest of
properties of the trajectory in the atmosphere were calculated for
each value of $d$.  In all cases the trajectories of the bodies were
integrated in the same gravitational scenario and using the same
parameters of the numerical integrator.

As a result of our Monte Carlo approach, a set of 50 different
reconstructed orbits were obtained.  The statistical properties of the
sample are presented in table \ref{tab:MonteCarlo}.

%TTTTTTTTTTTTTTTTTTTTTTTTTTTTTTTTTTTTTTTTTTTTTTTTTTTTTTTTTTTTTTTTTTTTTTTTTTT
%TABLE 2: MONTE CARLO ORBITS
\begin{table}[ht]
  \centering
  \begin{tabular}{llcccc}
    \hline\hline
    Property & Symbol (units) & Min. & Max. & Median & Mean $\pm$ St.Dev.\\\hline
    %%%%%%%%%%%%%%%%%%%%%%%%%%%%%%%%%%%%%%%%%%%%%%%%%%%%%%%%%%%%%%%%%%%%%%%%%%%%%%%%%
    Semimajor axis & $a$ (AU) & 1.40 & 2.21 & 1.69 & \bf 1.73 $\pm$ 0.23\\
    Eccentricity & e & 0.37 & 0.65 & 0.51 & \bf 0.51 $\pm$ 0.08\\
    Inclination & $i$ ($^o$) & 0.03 & 6.98 & 3.30 & \bf 3.45 $\pm$ 2.02\\
    Argument of periapsis & $\omega$ ($^o$) & 116.06 & 125.25 & 120.75 & \bf 120.62 $\pm$ 2.77\\
    Longitude of ascending node & $\Omega$ ($^o$) & 326.50 & 331.87 & 326.51 & \bf 326.70 $\pm$ 0.79\\
    Perihelion distance & $q$ (AU) & 0.77 & 0.88 & 0.82 & \bf 0.82 $\pm$ 0.03\\
    Aphelion distance & $Q$ (AU) & 1.93 & 3.64 & 2.55 & \bf 2.64 $\pm$ 0.49\\
    \hline\hline
  \end{tabular}
  \caption{Orbital elements statistics calculated for the Monte Carlo
    sample drawn in this work to reconstruct the orbit of the
    Chelyabinsk meteoroid.
  \label{tab:MonteCarlo}}
\end{table}
%TTTTTTTTTTTTTTTTTTTTTTTTTTTTTTTTTTTTTTTTTTTTTTTTTTTTTTTTTTTTTTTTTTTTTTTTTTT

As expected, a large uncertainty in the impact velocity leads to large
uncertainties in the orbital elements.  The most affected is the
semimajor axis.  Orbits with $a$ as large as 2.2 AU and as small as
1.4 AU were found varying the parameters of the meteoroid trajectory
in the atmosphere.  Despite the uncertainties, several bulk
characteristics of the orbit can be reliably established.
Eccentricity and perihelion distance are determined at a level of
around 10\% error or less. The inclination, although still uncertain
is in the order of magnitude of what we would expect.  Longitude of
the asscending node and argument of the periapsis are essentially
determined by the point in the meteoroid orbit where it intersects the
Earth's orbit.  Since all the orbits satisfy the condition to be in
the same position at the same date (the impact date) these two
elements are essentially the same across the whole sample.

\begin{figure}  
  \centering
   \includegraphics[width=1.0
  \textwidth]{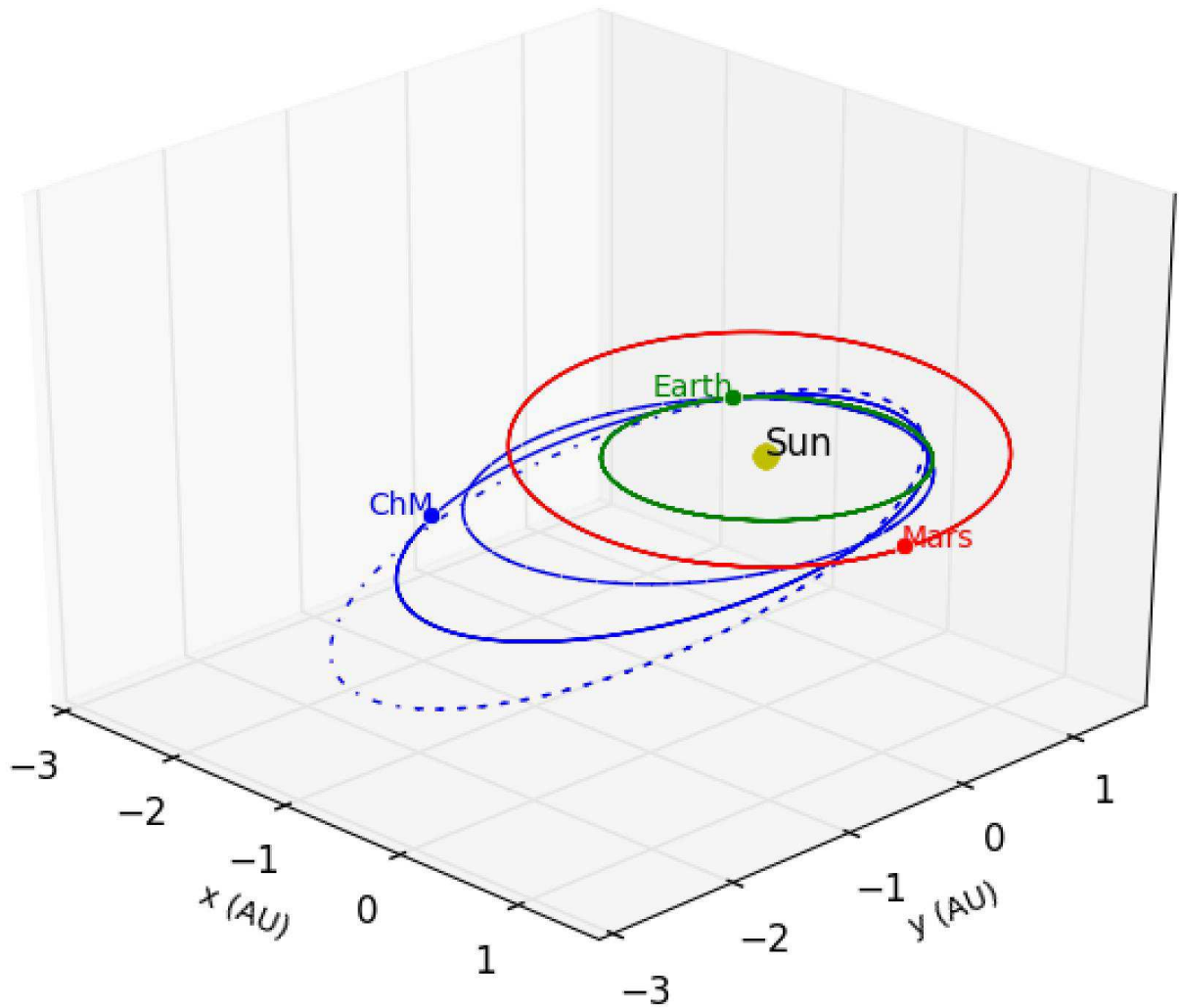}
  \scriptsize
  \caption{ Reconstructed orbits for the Chelyabinsk meteoroid.  For
    reference the orbits of Earth and Mars are represented.
    The rest of inner and outer planets though not shown were taken
    into account in the integration of the meteoroid orbit. The
    meteoroid orbit with the continuous line shows the orbit whose
    properties are closer to the median orbit, i.e. half of the
    orbits found in the Monte Carlo are bigger, more eccentric, and more
    inclined than this.  The dot-dashed and dashed orbits correspond
    to orbits 1-$sigma$ to the left and to the right of the median
    orbit.  
    \vspace{0.2cm}} 
  \label{fig:Orbits}
\end{figure}

To illustrate graphically this results we show in Figure
\ref{fig:Orbits} the reconstructed orbit and two extreme orbits
compatible with the uncertainties of our reconstruction.  We see in
that although we cannot individualize a given orbit, the general
features of the meteoroid trajectory were reconstructed successfully
with our procedure.

\begin{figure}  
  \centering
   \includegraphics[width=1.0
  \textwidth]{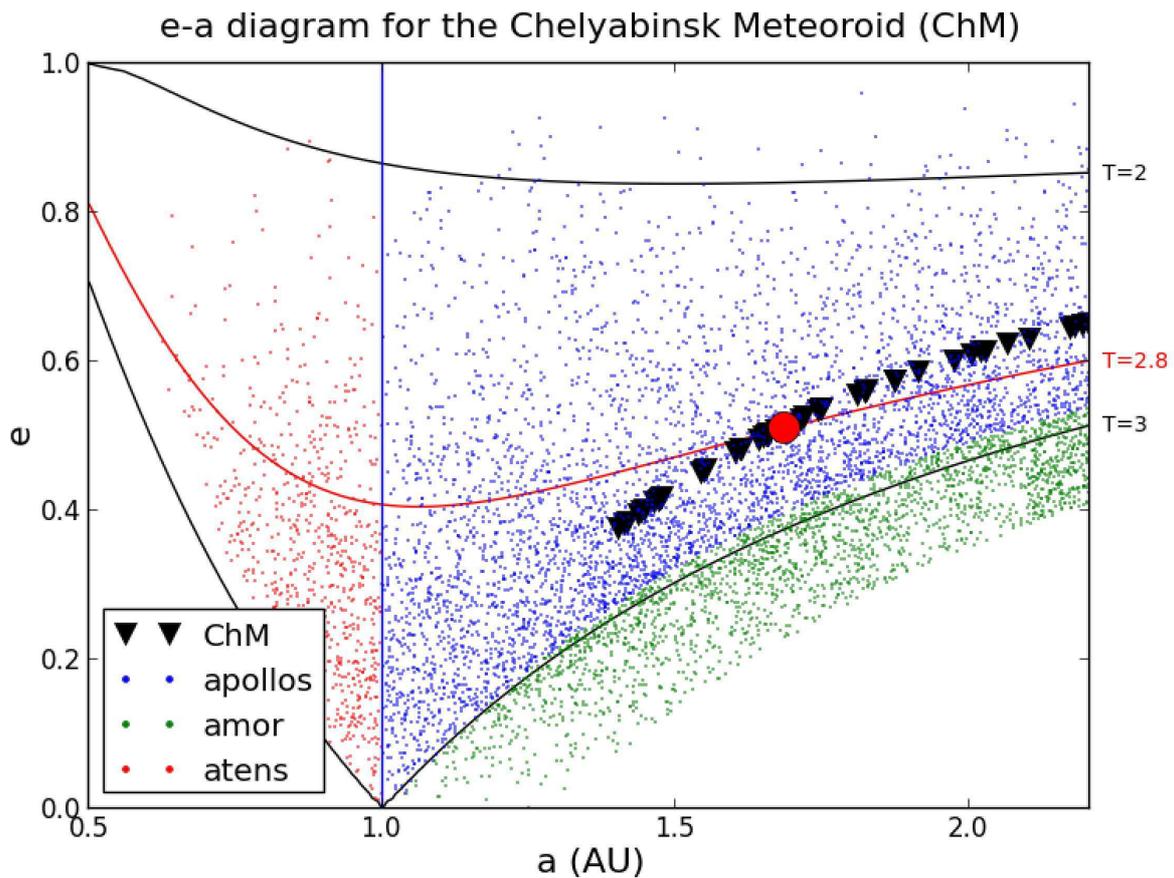}
  \scriptsize
  \caption{Orbital elements of the reconstructed orbits in the Monte
    Carlo sample (black triangles).  The red circle is close to the
    median values of $a$ and $e$. For comparison we included the
    orbital parameters of the Apollo asteroids (blue dots), Amor
    asteroids (green dots) and the Atens asteroids (red dots).  Curves
    of equal values of the Tisserand parameter relative to Earth,
    $T=a_E/a + 2\sqrt{a/a_E(1-e^2)}\cos i$ were also included for
    reference purposes.
    \vspace{0.2cm}} 
  \label{fig:a-e}
\end{figure}

Finally we want to classify the meteoroid by comparing its orbit with
that of already known asteroidal families.  To achieve this goal we
have plotted in an $a-e$ diagram the orbital elements in the random
sample and compare them with that of already known Asteroids belonging
to the Apollo, Amor and Atens families.  The results is depicted in
Figure \ref{fig:a-e}.  We can better appreciate here the large
uncertainties in semimajor axis arising from this type of
reconstruction procedure. According to this Figure the Chelyabinsk
meteoroid belonged inequivocally to the Apollo family of Asteroids.

%%%%%%%%%%%%%%%%%%%%%%%%%%%%%%%%%%%%%%%%%%%%%%%%%%%%%%%%%%%%%%%%%%%%%%%%%
\section{Discussion and Conclusions}
\label{sec:discussion}
%%%%%%%%%%%%%%%%%%%%%%%%%%%%%%%%%%%%%%%%%%%%%%%%%%%%%%%%%%%%%%%%%%%%%%%%%

We have reconstructed the orbit of the Chelyabinsk meteoroid.  We used
the most reliable information that can be found in the increasing
amount of evidence recorded in video in the affected area.  This is
not the first attempt at reconstructing the orbit of this important
object and will not be the final one.  However it is the most rigurous
reconstruction based solely in the evidence gathered in situ by
amateur and public cameras.

There are several important assumptions and hypothesis supporting our
results.  In the first place we assume that the time stamps used by
the camera at the Revolution Square in Cheliabinsk are precise.  Being
a public camera recorder it is expected that some care is put at
maintaining the clock of the camera on time.  This is not precisely
the case in most of the rest of videos available in the web where
camera clocks display times off by seconds to minutes with respect to
the time of the event.  We have verified by running a small Monte
Carlo sample of 10 orbits that the statistical conclusions presented
here are not affected when the initial time is shifted up to 5 seconds
with respect to that shown in the camera log.

The largest source of uncertainty in our reconstruction, is the
qualitative nature of the observational evidence in the second vantage
point used in the ``triangulation'' of the meteoroid trajectory.
Although every video filmed at Korkin shows that the fireball crossed
the small city skies almost vertically there are not quantitative
evidence supporting this or allowing us quantifying how ``vertical''
the phenomenon actually.  Further efforts to clarify this point should
be attempted to reduce the uncertainties in the direction and velocity
of the reconstructed trajectory.

Assuming that the hole in the ice sheet of Lake Cherbakul was produced
by a fragment of the meteoroid is also a very important hypothesis of
this work.  More importantly, our conclusions relies strongly onto
assume that the direction of the trajectory of the fragment
responsible for the breaking of the ice sheet in the Lake, is
essentially the same as the direction of the parent body.  It could be
not the case.  After the explosion and fragmentation of the meteoroid
fragments could acquire different velocities and fall affecting areas
far from the region wher we expect to find.

Our estimation of the meteoroid orbital velocity assume that this
quantity is constant during the penetration in the atmosphere.
Although velocities larger than 10 km/s should not be modified too
much before the meteoroid explosion or break up the points of the
trajectory used in our reconstruction were close to time of explosion
and fragmentation of the meteoroid.  Accordingly we would expect that
our estimations of the velocity will underestimates the orbital
velocity outside the atmosphee.  A consequence of this would be that
the size of the actual orbit could be larger than that obtained here.

% %%%%%%%%%%%%%%%%%%%%%%%%%%%%%%%%%%%%%%%%%%%%%%%%%%%%%%%%%%%%%%%%%%%%%%
% \section*{Acknowledgments}
% %%%%%%%%%%%%%%%%%%%%%%%%%%%%%%%%%%%%%%%%%%%%%%%%%%%%%%%%%%%%%%%%%%%%%%
\section*{Acknowledgments}
We appreciate the discussion with our colleagues at the institute of
Physics, Prof. Pablo Cuartas, Juan Carlos Muñoz and Carlos Molina.  We
thank to Mario Sucerquia by discovering and atracting our attention to
the blog of the Stefan Geen on which the method used in this work was
based.

%%%%%%%%%%%%%%%%%%%%%%%%%%%%%%%%%%%%%%%%%%%%%%%%%%%%%%%%%%%%%%%%%%%%%%%%%%%%%%%%%
%BIBLIOGRAPHY
%%%%%%%%%%%%%%%%%%%%%%%%%%%%%%%%%%%%%%%%%%%%%%%%%%%%%%%%%%%%%%%%%%%%%%%%%%%%%%%%%

\newpage

%%%%%%%%%%%%%%%%%%%%%%%%%%%%%%%%%%%%%%%%%%%%%%%%%%%%%%%%%%%%%%%%%%%%%%%%%%%%%%%%%
%FIGURES
%%%%%%%%%%%%%%%%%%%%%%%%%%%%%%%%%%%%%%%%%%%%%%%%%%%%%%%%%%%%%%%%%%%%%%%%%%%%%%%%%

%FFFFFFFFFFFFFFFFFFFFFFFFFFFFFFFFFFFFFFFFFFFFFFFFFFFFFFFFFFFFFFFFFFFFF
%FIGURE 1
%FFFFFFFFFFFFFFFFFFFFFFFFFFFFFFFFFFFFFFFFFFFFFFFFFFFFFFFFFFFFFFFFFFFFF
%FFFFFFFFFFFFFFFFFFFFFFFFFFFFFFFFFFFFFFFFFFFFFFFFFFFFFFFFFFFFFFFFFFFFF

%FFFFFFFFFFFFFFFFFFFFFFFFFFFFFFFFFFFFFFFFFFFFFFFFFFFFFFFFFFFFFFFFFFFFF
%FIGURE 2
%FFFFFFFFFFFFFFFFFFFFFFFFFFFFFFFFFFFFFFFFFFFFFFFFFFFFFFFFFFFFFFFFFFFFF
%FFFFFFFFFFFFFFFFFFFFFFFFFFFFFFFFFFFFFFFFFFFFFFFFFFFFFFFFFFFFFFFFFFFFF

%FFFFFFFFFFFFFFFFFFFFFFFFFFFFFFFFFFFFFFFFFFFFFFFFFFFFFFFFFFFFFFFFFFFFF
%FIGURE 3
%FFFFFFFFFFFFFFFFFFFFFFFFFFFFFFFFFFFFFFFFFFFFFFFFFFFFFFFFFFFFFFFFFFFFF
%FFFFFFFFFFFFFFFFFFFFFFFFFFFFFFFFFFFFFFFFFFFFFFFFFFFFFFFFFFFFFFFFFFFFF

\begin{thebibliography}{}

\bibitem[Acton~Jr(1996)Acton~Jr]{Acton1996}
Acton~Jr, C.~H. 1996.
\newblock Ancillary data services of nasa's navigation and ancillary
  information facility.
\newblock {\em Planetary and Space Science\/}~{\em 44\/}(1), 65--70.

\bibitem[Bangert {\em et~al.}(2011)Bangert, Puatua, Kaplan, Bartlett, Harris,
  Fredericks, and Monet]{Bangert2011}
Bangert, J., W.~Puatua, G.~Kaplan, J.~Bartlett, W.~Harris, A.~Fredericks,\ and
  A.~Monet 2011.
\newblock User’s guide to novas version c3.1.

\bibitem[Chambers(2008)Chambers]{Chambers2008}
Chambers, J. 2008.
\newblock A hybrid symplectic integrator that permits close encounters between
  massive bodies.
\newblock {\em Monthly Notices of the Royal Astronomical Society\/}~{\em
  304\/}(4), 793--799.

\bibitem[Fischer {\em et~al.}(2012)Fischer, Schwamb, Schawinski, Lintott,
  Brewer, Giguere, Lynn, Parrish, Sartori, Simpson, et~al.]{Fischer2012}
Fischer, D.~A., M.~E. Schwamb, K.~Schawinski, C.~Lintott, J.~Brewer,
  M.~Giguere, S.~Lynn, M.~Parrish, T.~Sartori, R.~Simpson, et~al. 2012.
\newblock Planet hunters: the first two planet candidates identified by the
  public using the kepler public archive data★.
\newblock {\em Monthly Notices of the Royal Astronomical Society\/}.

\bibitem[Lintott {\em et~al.}(2008)Lintott, Schawinski, Slosar, Land, Bamford,
  Thomas, Raddick, Nichol, Szalay, Andreescu, et~al.]{Lintott2008}
Lintott, C.~J., K.~Schawinski, A.~Slosar, K.~Land, S.~Bamford, D.~Thomas, M.~J.
  Raddick, R.~C. Nichol, A.~Szalay, D.~Andreescu, et~al. 2008.
\newblock Galaxy zoo: morphologies derived from visual inspection of galaxies
  from the sloan digital sky survey★.
\newblock {\em Monthly Notices of the Royal Astronomical Society\/}~{\em
  389\/}(3), 1179--1189.

\bibitem[Trigo-Rodr{\'\i}guez {\em et~al.}(2009)Trigo-Rodr{\'\i}guez, Madiedo,
  Williams, Castro-Tirado, Llorca, V{\'\i}tek, and Jel{\'\i}nek]{Trigo2009}
Trigo-Rodr{\'\i}guez, J.~M., J.~M. Madiedo, I.~P. Williams, A.~J.
  Castro-Tirado, J.~Llorca, S.~V{\'\i}tek,\ and M.~Jel{\'\i}nek 2009.
\newblock Observations of a very bright fireball and its likely link with comet
  c/1919 q2 metcalf.
\newblock {\em Monthly Notices of the Royal Astronomical Society\/}~{\em
  394\/}(1), 569--576.

\end{thebibliography}
\end{document}